\documentclass{sig-alternate-05-2015}

\usepackage{graphicx} 
\usepackage{subfigure} 
\usepackage[utf8]{inputenc}

\usepackage{algorithm}
\usepackage{algorithmic}
\usepackage{amssymb}
\usepackage{amsmath}
\usepackage{bm}
\usepackage{mathtools}
\usepackage{paralist}
\usepackage{bbm}

\usepackage{etoolbox}
\newtoggle{ARXIV}
\toggletrue{ARXIV}

\usepackage{booktabs}
\usepackage[pdfpagelabels=false]{hyperref}
\usepackage[usenames,dvipsnames]{xcolor}
\usepackage{pgf}

\DeclareMathOperator{\rel}{rel}
\DeclareMathOperator{\relx}{r}
\DeclareMathOperator{\click}{c}
\DeclareMathOperator{\exam}{e}
\DeclareMathOperator{\rank}{rank}

\DeclareMathOperator{\argmin}{argmin}

\DeclareMathOperator{\Prob}{P}
\newcommand{\x}{\bm{x}}                  
\newcommand{\xsample}{\bm{X}}            
\newcommand{\y}{\bm{y}}                  
\newcommand{\ypres}{\bar{\bm{y}}}        
\newcommand{\ypart}{y}                   
\newcommand{\obs}{o}                     
\newcommand{\Loss}{\Delta}               
\newcommand{\Risk}{R}                    
\newcommand{\Losshat}{\hat{\Loss}}       
\newcommand{\Riskhat}{\hat{\Risk}}       

\newcommand{\E}{{\mathbb{E}}}            
\newcommand{\systems}{{\mathcal{S}}}     
\newcommand{\system}{{S}}                
\newcommand{\n}{{n}}                     
\newcommand{\N}{{N}}                     
\newcommand{\w}{w}                       
\newcommand{\what}{\hat{w}}              

\begin{document}


\doi{10.475/123_4}
\isbn{123-4567-24-567/08/06}
\iftoggle{ARXIV}{
\conferenceinfo{Under review}{}
}{
\conferenceinfo{WSDM '17}{February 6--10, 2017, Cambridge, UK}
}
\acmPrice{\$15.00}

%
\title{Unbiased Learning-to-Rank with Biased Feedback}
\iftoggle{ARXIV}{
\numberofauthors{3} 
\author{
\alignauthor Thorsten Joachims\\
       \affaddr{Cornell University, Ithaca, NY}\\
       \email{tj@cs.cornell.edu}
\alignauthor
Adith Swaminathan\\
       \affaddr{Cornell University, Ithaca, NY}\\
       \email{adith@cs.cornell.edu}
\alignauthor
Tobias Schnabel\\
       \affaddr{Cornell University, Ithaca, NY}\\
       \email{tbs49@cornell.edu}
}
}{
\numberofauthors{1} 
\author{
<omitted>
}
}

\maketitle
\begin{abstract}
Implicit feedback (e.g., clicks, dwell times, etc.) is an abundant source of data in human-interactive systems. While implicit feedback has many advantages (e.g., it is inexpensive to collect, user centric, and timely), its inherent biases are a key obstacle to its effective use. For example, position bias in search rankings strongly influences how many clicks a result receives, so that directly using click data as a training signal in Learning-to-Rank (LTR) methods yields 
sub-optimal results. To overcome this bias problem, we present a counterfactual inference framework that provides the theoretical basis for unbiased LTR via Empirical Risk Minimization despite biased data.
Using this framework, we derive a Propensity-Weighted Ranking SVM for discriminative learning from implicit feedback, where click models take the role of the propensity estimator. In contrast to most conventional approaches to de-bias the data using click models, this allows training of ranking functions even in settings where queries do not repeat. Beyond the theoretical support, we show empirically that the proposed learning method is highly effective in dealing with biases, that it is robust to noise and propensity model misspecification, and that it scales efficiently. We also demonstrate the real-world applicability of our approach on an operational search engine, where it substantially improves retrieval performance.
\end{abstract}

%
%

%
%

%
%


\section{Introduction}
\label{sec:intro}
Batch training of retrieval systems requires annotated test collections that take substantial effort and cost to amass. While economically feasible for Web Search, eliciting relevance annotations from experts is infeasible or impossible for most other ranking applications (e.g., personal collection search, intranet search). For these applications, implicit feedback from user behavior is an attractive source of data. Unfortunately, existing approaches for Learning-to-Rank (LTR) from implicit feedback -- and clicks on search results in particular -- have several limitations or drawbacks.

First, the na\"{\i}ve approach of treating a click/no-click as a positive/negative relevance judgment is severely biased. In particular, the order of presentation has a strong influence on where users click \cite{Joachims/etal/07a}.  This presentation bias leads to an incomplete and skewed sample of relevance judgments that is far from uniform, thus leading to biased learning-to-rank.

Second, treating clicks as preferences between clicked and skipped documents has been found to be accurate \cite{Joachims/02c,Joachims/etal/07a}, but it can only infer preferences that oppose the presented order. This again leads to severely biased data, and learning algorithms trained with these preferences tend to reverse the presented order unless additional heuristics are used \cite{Joachims/02c}. 

Third, probabilistic click models (see \cite{Chuklin/etal/15a}) have been used to model how users produce clicks, and they can take position and context biases into account. By estimating latent parameters of these generative click models, one can infer the relevance of a given document for a given query. However, inferring reliable relevance judgments typically requires that the same query is seen multiple times, which is unrealistic in many retrieval settings (e.g., personal collection search) and for tail queries.

Fourth, allowing the LTR algorithm to randomize what is presented to the user, like in online learning algorithms \cite{Raman/etal/13a,hofmann2013reusing} and batch learning from bandit feedback (BLBF) \cite{Swaminathan/Joachims/15c} can overcome the problem of bias in click data in a principled manner. However, requiring that rankings be actively perturbed during system operation whenever we collect training data decreases ranking quality and, therefore, incurs a cost compared to observational data collection. 

In this paper we present a theoretically principled and empirically effective approach for learning from observational implicit feedback that can overcome the limitations outlined above. By drawing on counterfactual estimation techniques from causal inference \cite{Imbens/Rubin/15}, we first develop a provably unbiased estimator for evaluating ranking performance using biased feedback data. Based on this estimator, we propose a Propensity-Weighted Empirical Risk Minimization (ERM) approach to LTR, which we implement efficiently in a new learning method we call Propensity SVM-Rank. While our approach uses a click model, the click model is merely used to assign propensities to clicked results in hindsight, not to extract aggregate relevance judgments. This means that our Propensity SVM-Rank does not require queries to repeat, making it applicable to a large range of ranking scenarios. 
Finally, our methods can use observational data and we do not require that the system randomizes rankings during data collection, except for a small pilot experiment to estimate the propensity model.

When deriving our approach, we provide theoretical justification for each step, leading to a rigorous end-to-end approach that does not make unspecified assumptions or employs heuristics. This provides a principled basis for further improving components of the approach (e.g., the click propensity model, the ranking performance measure, the learning algorithm). We present an extensive empirical evaluation testing the limits of the approach on synthetic click data, finding that it performs robustly over a large range of bias, noise, and misspecification levels. Furthermore, we field our method in a real-world application on an operational search engine, finding that it is robust in practice and manages to substantially improve retrieval performance. 


\section{Related Work}
\label{sec:related}
There are two groups of approaches for handling biases in implicit feedback for learning-to-rank. The first group assumes the feedback collection step is fixed, and tries to interpret the observationally collected data so as to minimize bias effects. Approaches in the second group intervene during feedback collection, trying to present rankings that will lead to less biased feedback data overall.   

Approaches in the first group commonly assume some model of user behavior in order to explain bias effects. For example, in a cascade model \cite{craswell2008position}, users are assumed to sequentially go down a ranking and click on a document if it is relevant.
Clicks, under this model, let us learn preferences between skipped and clicked documents. Learning from these relative preferences lowers the impact of some biases \cite{Joachims/02c}. 
Other 
click models (\cite{craswell2008position,Chapelle/Zhang/09,borisov2016neural}, also see \cite{Chuklin/etal/15a}) 
have been proposed, and are trained to maximize log-likelihood of observed clicks.
In these click modeling approaches, performance on downstream learning-to-rank algorithms is merely an afterthought. 
In contrast, we separate click propensity estimation and learning-to-rank in a principled way and we optimize for ranking performance directly. 
Our framework allows us to plug-and-play more sophisticated user models in place of the simple click models we use in this work.  


The key technique used by approaches in the second group to obtain more reliable click data are randomized experiments. For instance, randomizing documents across all ranks lets us learn unbiased relevances for each document, and swapping neighboring pairs of documents \cite{Raman/Joachims/13a} lets us learn reliable pairwise preferences.
Similarly, randomized interleaving can detect preferences between different rankers reliably \cite{Chapelle/etal/12a}. 
Different from online learning via bandit algorithms and interleaving \cite{Yue/Joachims/09a,schuth2016multileave}, batch learning from bandit feedback (BLBF) \cite{Swaminathan/Joachims/15c}
still uses randomization during feedback collection, and then performs offline learning. Our problem formulation can be interpreted as being half way between the BLBF setting (loss function is unknown and no assumptions on loss function) and learning-to-rank from editorial judgments (components of ranking are fully labeled and loss function is given) since we know the form of the loss function but labels for only some parts of the ranking are revealed.
All approaches that use randomization suffer from two limitations. First, randomization typically degrades ranking quality during data collection; second, deploying non-deterministic ranking functions introduces bookkeeping overhead.  
In this paper, the system can be deterministic and we merely exploit and model stochasticity in user behavior. Moreover, our framework also allows (but does not require) the use of randomized data collection in order to mitigate the effect of biases and improve learning. 

Our approach uses inverse propensity scoring (IPS), originally employed in causal inference from observational studies \cite{Rosenbaum1983}, and more recently also in whole page optimization \cite{Wang2016}, IR evaluation with manual judgments \cite{Schnabel/etal/16c}, and recommender evaluation \cite{li2011unbiased,Schnabel/etal/16b}. 
We use randomized interventions similar to \cite{craswell2008position,Strehl2010,Wang/etal/16} to estimate propensities in a position discount model. Unlike the uniform ranking randomization of \cite{Wang/etal/16} (with its high performance impact) or swapping adjacent pairs as in \cite{craswell2008position}, we swap documents in different ranks to the top position randomly as in \cite{Strehl2010}. See Section~\ref{sec:prop_est} for details.

Finally, our approach is similar in spirit to  \cite{Wang/etal/16}, where propensity-weighting is used to correct for selection bias when discarding queries without clicks during learning-to-rank. 
The key insight of our work is to recognize that
inverse propensity scoring can be employed much more powerfully, to account for position bias, trust bias, contextual effects, document popularity etc.\ using appropriate click models to estimate the propensity of each click rather than the propensity for a query to receive a click as in \cite{Wang/etal/16}.



\section{Full-Info Learning to Rank}

Before we derive our approach for LTR from biased implicit feedback, we first review the conventional problem of LTR from editorial judgments. In conventional LTR, 
we are given a sample $\xsample$ of i.i.d.\ queries $\x_i \sim \Prob(\x)$ for which we assume the relevances $\rel(\x,\ypart)$ of all documents $\ypart$ are known. Since all relevances are assumed to be known, 
we call this the Full-Information Setting. The relevances can be used to compute the {\em loss} $\Loss(\y|\x)$ (e.g., negative DCG) of any ranking $\y$ for query $\x$. Aggregating the losses of individual rankings by taking the expectation over the query distribution, we can define the overall {\em risk} 
of a ranking system $\system$ that returns rankings $\system(\x)$ as
\begin{eqnarray}
    \Risk(\system) & = & \int \Loss(\system(\x)|\x) \: d \Prob(\x). \label{eq:utilfullinfo}
\end{eqnarray}
The goal of learning is to find a ranking function $\system \in \systems$ that minimizes $\Risk(\system)$ for the query distribution $\Prob(\x)$. Since $\Risk(\system)$ cannot be computed directly, it is typically estimated via the {\em empirical risk}
\begin{eqnarray}
    \Riskhat(\system) & = & \frac{1}{|\xsample|}\sum_{\x_i \in \xsample} \Loss(\system(\x_i)|\x_i). \nonumber
\end{eqnarray}
A common learning strategy is {\em Empirical Risk Minimization (ERM)} \cite{Vapnik1998}, which corresponds to picking the system $\hat{\system} \in \systems$ that optimizes the empirical risk
\begin{eqnarray}
    \hat{\system} & = & \argmin_{\system \in \systems} \left\{ \Riskhat(\system) \right\}, \nonumber
\end{eqnarray}
possibly subject to some regularization in order to control overfitting. There are several LTR algorithms that follow this approach (see \cite{Liu2009}), and we use SVM-Rank \cite{Joachims/02c} as a representative algorithm in this paper.

The relevances $\rel(\x,\ypart)$ are typically elicited via expert judgments. Apart from being expensive and often infeasible (e.g., in personal collection search), expert judgments come with at least two other limitations. First, since it is clearly impossible to get explicit judgments for all documents, pooling techniques \cite{SparckJones1975} are used such that only the most promising documents are judged. While cutting down on judging effort, this introduces an undesired \emph{pooling bias} because all unjudged documents are typically assumed to be irrelevant. 
The second limitation is that expert judgments $\rel(\x,\ypart)$ have to be aggregated over all intents that underlie the same query string, and it can be challenging for a judge to properly conjecture the distribution of query intents to assign an appropriate $\rel(\x,\ypart)$.

\section{Partial-Info Learning to Rank} \label{sec:partial_ltr}

Learning from implicit feedback has the potential to overcome the above-mentioned limitations of full-information LTR. By drawing the training signal directly from the user, it naturally reflects the user's intent, since each user acts upon their own relevance judgement subject to their specific context and information need. It is therefore more appropriate to talk about query instances $\x_i$ that include contextual information about the user, instead of query strings $\x$. For a given query instance $\x_i$, we denote with $\relx_{i}(\ypart)$ the user-specific relevance of result $\ypart$ for query instance $\x_i$. One may argue that what expert assessors try to capture with $\rel(\x,\ypart)$ is the mean of the relevances $\relx_{i}(\ypart)$ over all query instances that share the query string, so,
using implicit feedback for learning is able to remove a lot of guesswork about what the distribution of users meant by a query.

However, when using implicit feedback as a relevance signal, unobserved feedback is an even greater problem than missing judgments in the pooling setting. In particular, implicit feedback is distorted by presentation bias, and it is not missing completely at random \cite{rubin2002mnar}. To nevertheless derive well-founded learning algorithms, we adopt the following counterfactual model. It closely follows \cite{Schnabel/etal/16c}, which unifies several prior works on evaluating information retrieval systems. 

For concreteness and simplicity, assume that relevances are binary, $\relx_i(\ypart) \in \{0,1\}$, and our performance measure of interest is the sum of the ranks of the relevant results
\begin{eqnarray}
    \Loss(\y|\x_i,\relx_i) & = & \sum_{\ypart \in \y} \rank(\ypart|\y) \cdot \relx_i(\ypart). \label{eq:avgrankfullinfo}
\end{eqnarray}
Analogous to \eqref{eq:utilfullinfo}, we can define the risk of a system as 
\begin{eqnarray}
    \Risk(\system) & = & \int \Loss(\system(\x)|\x,\relx) \: d \Prob(\x,\relx). \label{eq:utilpartialinfo}
\end{eqnarray}
In our counterfactual model, there exists a true vector of relevances $\relx_i$ for each incoming query instance $(\x_i,\relx_i) \sim \Prob(\x,\relx)$. However, only a part of these relevances is observed for each query instance, while typically most remain unobserved. In particular, given a presented ranking $\ypres_i$ we are more likely to observe the relevance signals (e.g., clicks) for the top-ranked results than for results ranked lower in the list. Let $\obs_i$ denote the 0/1 vector indicating which relevance values were revealed, $\obs_i \sim \Prob(\obs|\x_i, \ypres_i,\relx_i)$. 
For each element of $\obs_i$, denote with $Q(\obs_i(\ypart)=1|\x_i,\ypres_i, \relx_i)$ the marginal probability of observing the relevance $\relx_i(\ypart)$ of result $\ypart$ for query $\x_i$, if the user was presented the ranking $\ypres_i$. We refer to this probability value as the \emph{propensity} of the observation.
We will discuss how $\obs_i$ and $Q$ can be obtained in Section~\ref{sec:feedbackmodels}.
 
Using this counterfactual modeling setup, we can get an unbiased estimate of $\Loss(\y|\x_i,\relx_i)$ for any new ranking $\y$ (typically different from the presented ranking $\ypres_i$) via the inverse propensity scoring (IPS) estimator \cite{Horvitz1952,Rosenbaum1983,Imbens/Rubin/15}
\begin{eqnarray}
    \Losshat_{IPS}(\y|\x_i,\ypres_i, \obs_i) & \!\!\!=\!\!\! & \sum_{\ypart : \obs_i(\ypart) = 1}  \frac{\rank(\ypart|\y) \!\cdot\! \relx_i(\ypart)}{Q(\obs_{i}(\ypart)\!=\!1|\x_i,\ypres_i,\relx_i)} \nonumber \\
    & \!\!\!=\!\!\! & \sum_{\substack{ \ypart : \obs_i(\ypart) = 1 \\
    \bigwedge \relx_i(\ypart) = 1}}  \frac{\rank(\ypart|\y) }{Q(\obs_{i}(\ypart)\!=\!1|\x_i,\ypres_i,\relx_i)}. \nonumber
\end{eqnarray}
This is an unbiased estimate of $\Loss(\y|\x_i,\relx_i)$ for any $\y$, if $Q(\obs_i(\ypart)=1|\x_i,\ypres_i,\relx_i) > 0$ for all $\ypart$ that are relevant $\relx_i(\ypart) = 1$ (but not necessarily for the irrelevant $\ypart$). \begin{align*}
    \!\!\!\!\!\!\!\!\!\!\!\!\!\!\!\!\!\!\!\!\!\!\!\!\!\!\!\!\!\!\!\!\!\!\!\!\!\!\!\!\!\!\E_{\obs_i}[\Losshat_{IPS}(\y|\x_i,\ypres_i, \obs_i)] & & 
\end{align*} \vspace*{-0.6cm}
\begin{eqnarray}
    & = & \E_{\obs_i}\!\!\left[ \sum_{\ypart : \obs_i(\ypart) = 1}  \frac{\rank(\ypart|\y) \!\cdot\! \relx_i(\ypart)}{Q(\obs_{i}(\ypart)\!=\!1|\x_i,\ypres_i,\relx_i)} \right] \nonumber \\
    & = & \sum_{\ypart \in \y} \E_{\obs_i}\!\!\left[  \frac{\obs_i(\ypart) \!\cdot\! \rank(\ypart|\y) \!\cdot\! \relx_i(\ypart)}{Q(\obs_{i}(\ypart)\!=\!1|\x_i,\ypres_i,\relx_i))} \right] \nonumber \\
    & = & \sum_{\ypart \in \y} \frac{Q(\obs_i(\ypart)=1|\x_i,\ypres_i,\relx_i) \cdot \rank(\ypart|\y) \cdot \relx_i(\ypart) }{Q(\obs_i(\ypart)=1|\x_i,\ypres_i,\relx_i)}  \nonumber \\
    & = & \sum_{\ypart \in \y} \rank(\ypart|\y) \relx_i(\ypart)  \nonumber \\
    & = & \Loss(\y|\x_i,\relx_i) .\nonumber
\end{eqnarray}
The second step uses linearity of expectation, and the fourth step uses $Q(\obs_i(\ypart)=1|\x_i,\ypres_i,\relx_i) > 0$. 

An interesting property of $\Losshat_{IPS}(\y|\x_i,\ypres_i, \obs_i)$ is that only those results $\ypart$ with $[\obs_i(\ypart)=1 \wedge \relx_i(\ypart)=1]$ (i.e. clicked results, as we will see later) contribute to the estimate. 
We therefore only need the propensities $Q(\obs_i(\ypart)=1|\x_i,\ypres_i,\relx_i)$ for relevant results.
Since we will eventually need to estimate the propensities $Q(\obs_i(\ypart)=1|\x_i,\ypres_i,\relx_i)$, an additional requirement for making $\Losshat_{IPS}(\y|\x_i,\ypres_i,\obs_i)$ computable while remaining unbiased is that the propensities only depend on observable information (i.e., unconfoundedness, see \cite{Imbens/Rubin/15}).

To define the empirical risk to optimize during learning, we begin by collecting a sample of $\N$ query instances $\x_i$, recording the partially-revealed relevances $\relx_i$ as indicated by $\obs_i$, and the propensities $Q(\obs_{i}(\ypart)=1|\x_i,\ypres_i,\relx_i)$ for the observed relevant results in the ranking $\ypres_i$ presented by the system.
Then, the empirical risk of a system is simply the IPS estimates averaged over query instances:
\begin{eqnarray}
    \!\!\!\!\Riskhat_{IPS}(\system) \!\!\!\!& = \!\!\!\!& \frac{1}{\N}\!\sum_{i=1}^{\N} \sum_{\substack{ \ypart : \obs_i(\ypart) = 1 \\
    \bigwedge \relx_i(\ypart) = 1}}  \frac{\rank(\ypart|\system(\x_i)) }{Q(\obs_{i}(\ypart)\!=\!1|\x_i,\ypres_i,\relx_i)} . \:\:\:\: \label{eq:riskips}
\end{eqnarray}
Since $\Losshat_{IPS}(\y|\x_i,\ypres_i,\obs_i)$ is unbiased for each query instance, the aggregate $\Riskhat_{IPS}(\system)$ is also unbiased for $\Risk(\system)$ from \eqref{eq:utilpartialinfo},
\begin{eqnarray}
    \E[\Riskhat_{IPS}(\system)] & = & \Risk(\system). \nonumber
\end{eqnarray}  
Furthermore, it is easy to verify that $\Riskhat_{IPS}(\system)$ converges to the true $\Risk(\system)$ under mild additional conditions (i.e., propensities bounded away from $0$) as we increase the sample size $\N$ of query instances. 
So, we can perform ERM using this propensity-weighted empirical risk,
\begin{eqnarray}
    \hat{\system} & = & \argmin_{\system \in \systems} \left\{\Riskhat_{IPS}(\system) \right\}. \nonumber
\end{eqnarray}
 Finally, using standard results from statistical learning theory \cite{Vapnik1998}, consistency of the empirical risk paired with capacity control implies consistency also for ERM. 
 In intuitive terms, this means that given enough training data, the learning algorithm is guaranteed to find the best system in 
 $\systems$.


\section{Feedback Propensity Models} \label{sec:feedbackmodels}

In Section~\ref{sec:partial_ltr}, we showed that the relevance signal $\relx_i$, the observation pattern $\obs_i$, and the propensities of the observations $Q(\obs_i(\ypart)=1|\x_i,\ypres_i,\relx_i)$ are the key components for unbiased LTR from biased observational feedback. We now outline how these quantities can be elicited and modeled in a typical search-engine application. However, the general framework of Section~\ref{sec:partial_ltr} extends beyond this particular application, and beyond the particular feedback model below.

\subsection{Position-Based Propensity Model}
\label{sec:position_feedbackmodel}

Search engine click logs provide
a sample of query instances $\x_i$,
the presented ranking $\ypres_i$ and a (sparse) click-vector where each $\click_i(y) \in \{0,1\}$ indicates
whether result $\ypart$ was clicked or not.
To derive propensities of observed clicks, we will employ 
a click propensity model.
For simplicity, we consider a straightforward examination model analogous to \cite{Richardson2007}, where a click on a search result depends on the probability that a user examines a result (i.e., $\exam_i(\ypart)$) and then decides to click on it (i.e., $\click_i(\ypart)$) in the following way:
\begin{eqnarray}
    P(\exam_i(\ypart)=1|\rank(\ypart|\ypres)) \cdot P(\click_i(\ypart)=1|\relx_i(\ypart),\exam_i(\ypart)=1) . \nonumber 
\end{eqnarray}
In this model, examination depends only on the rank of $\ypart$ in $\ypres$. So, $P(\exam_i(\ypart)=1|\rank(\ypart|\ypres_i))$ can be represented by a vector of examination probabilities $p_r$, one for each rank $r$. These examination probabilities can model presentation bias documented in eye-tracking studies \cite{Joachims/etal/07a}, where users are more likely to see results at the top of the ranking than those further down.

For the probability of click on an examined result $P(\click_i(\ypart)=1|\relx_i(\ypart),\exam_i(\ypart)=1)$, we first consider the simplest model where clicking is a deterministic noise-free function of the users private relevance assessment $\relx_i(\ypart)$. Under this model, users click if and only if the result is examined and relevant ($\click_i(\ypart)=1 \leftrightarrow [\exam_i(\ypart)=1 \: \wedge \: \relx_i(\ypart)=1]$). This means that for examined results (i.e., $\exam_i(\ypart)=1$) clicking is synonymous with relevance ($\exam_i(\ypart)=1 \rightarrow [\click_i(\ypart) = \relx_i(\ypart)]$). Furthermore, it means that we observe the value of $\relx_i(\ypart)$ perfectly when 
$\exam_i(\ypart)=1$ ($\exam_i(\ypart)=1 \rightarrow \obs_i(\ypart)=1$), and that we gain no knowledge of the true $\relx_i(\ypart)$ when a result is not examined ($\exam_i(\ypart)=0 \rightarrow \obs_i(\ypart)=0$). Therefore, examination equals observation and $Q(\obs_i(\ypart)|\x_i,\ypres_i,\relx_i) \equiv P(\exam_i(\ypart)|\rank(\ypart|\ypres_i))$. 

Using these equivalences, we can simplify the IPS estimator from \eqref{eq:riskips} by substituting $p_r$ as the propensities and by using $\click_i(\ypart)=1 \leftrightarrow [\obs_i(\ypart)=1 \: \wedge \: \relx_i(\ypart)=1]$
\begin{eqnarray}
    \Riskhat_{IPS}(\system) \!\!\!\!& = \!\!\!\!& \frac{1}{\n}\sum_{i=1}^{\n} \sum_{\ypart : \click_i(\ypart) = 1 }  \frac{\rank(\ypart|\system(\x_i)) }{p_{\rank(\ypart|\ypres_i)}} . \label{eq:riskips_posmodel}
\end{eqnarray}
$\Riskhat_{IPS}(\system)$ is an unbiased estimate of $\Risk(\system)$ under the position-based propensity model if $p_r > 0$ for all ranks.
While absence of a click does not imply that the result is not relevant (i.e., $\click_i(\ypart)=0 \not\rightarrow \relx_i(\ypart)=0$), the IPS estimator has the nice property that such explicit negative judgments are not needed to compute an unbiased estimate of $\Risk(\system)$ for the loss in \eqref{eq:avgrankfullinfo}.
Similarly, while absence of a click leaves us unsure about whether the result was examined (i.e., $\exam_i(\ypart)=?$), the IPS estimator only needs to know the indicators $\obs_i(\ypart)=1$ for results that are also relevant (i.e., clicked results).

Finally, note the conceptual difference in how we use this standard examination model compared to most prior work. We do not try to estimate an average relevance rating $\rel(\x,\ypart)$ by taking repeat instances of the same query $\x$, but we use the model as a propensity estimator to de-bias individual observed user judgments $\relx_i(\ypart)$ to be used directly in ERM.

\subsection{Incorporating Click Noise}
\label{sec:noisy_clicks}
In Section~\ref{sec:position_feedbackmodel}, we assumed that clicks reveal the user's true $\relx_i$ in a noise-free way. This is clearly unrealistic. In addition to the stochasticity in the examination distribution $P(\exam_i(\ypart)=1|\rank(\ypart|\y))$, we now also consider noise in the distribution that generates the clicks. In particular, we no longer require that a relevant result is clicked with probability $1$ and an irrelevant result is clicked with probability $0$, but instead, for $1 \ge \epsilon_+ > \epsilon_- \ge 0$,
\begin{eqnarray}
    P(\click_i(\ypart)=1|\relx_i(\ypart)=1,\obs_i(\ypart)=1)=\epsilon_+,  \nonumber \\
    P(\click_i(\ypart)=1|\relx_i(\ypart)=0,\obs_i(\ypart)=1)=\epsilon_-.  \nonumber
\end{eqnarray}
The first line means that users click on a relevant result only with probability $\epsilon_+$, while the second line means that users may erroneously click on an irrelevant result with probability $\epsilon_-$. An alternative and equivalent way of thinking about click noise is that users still click deterministically as in the previous section, but based on a noisily corrupted version $\tilde{\relx}_i$ of $\relx_i$. This means that all reasoning regarding observation (examination) events $\obs_i$ and their propensities $p_r$ still holds, and that we still have that $\click_i(\ypart)=1 \rightarrow \obs_i(\ypart)=1$. What does change, though, is that we no longer observe the ``correct'' $\relx_i(\ypart)$ but instead get feedback according to the noise-corrupted version $\tilde{\relx}_i(\ypart)$.
What happens to our learning process if we 
estimate risk using \eqref{eq:riskips_posmodel}, but now with $\tilde{\relx}_i$?

Fortunately, the noise does not affect ERM's ability to find the best ranking system given enough data.
While using noisy clicks leads to biased empirical risk estimates w.r.t. the true $\relx_i$ 
(i.e., $\E[\Riskhat_{IPS}(\system)] \not= \Risk(\system)$), in expectation this bias is order preserving for $\Risk(\system)$ such that the risk minimizer remains the same.
\begin{eqnarray*}
   \!\!& & \!\!\!\E[\Riskhat_{IPS}(\system_1)] > \E[\Riskhat_{IPS}(\system_2)] \\
   \!\!& \Leftrightarrow & \!\!\!\E_{\x, \relx, \ypres} \!\!\left[ \!\E_{\obs}  \E_{\click \!|\! \obs} \!\!\left[ \! \sum_{\ypart : \click(\ypart) = 1 } \!\!\!\!\! \frac{\rank(\ypart|\system_1\!(\!\x\!)\!) \!-\! \rank(\ypart|\system_2(\!\x\!)\!) }{p_{\rank(\ypart|\ypres)}} \right] \!\right] \!\! > \! 0 \\
   \!\!& \Leftrightarrow & \!\!\!\E_{\x, \relx} \!\!\left[  \sum_{\ypart} \Prob(\click(\ypart)=1|\obs(\ypart)=1,\relx(\ypart)) \delta\rank(\ypart|\x) \right]  > 0 \\
   \!\!& \Leftrightarrow & \!\!\!\E_{\x, \relx} \!\!\left[  \sum_{\ypart} \delta\rank(\ypart|\x) \cdot (\epsilon_+ \relx(\ypart) + \epsilon_-(1 - \relx(\ypart))) \right] > 0 \\
   \!\!& \Leftrightarrow & \!\!\!\E_{\x, \relx} \!\!\left[  \sum_{\ypart} \delta\rank(\ypart|\x) \cdot ((\epsilon_+ - \epsilon_-) \relx(\ypart) + \epsilon_-) \right] > 0 \\
  *\!\!\!\!\!& \Leftrightarrow & \!\!\!\E_{\x, \relx} \!\!\left[  \sum_{\ypart} \delta\rank(\ypart|\x) \cdot (\epsilon_+ - \epsilon_-) \relx(\ypart)  \right] > 0 \\
   \!\!& \Leftrightarrow & \!\!\!\E_{\x, \relx} \!\!\left[  \sum_{\ypart} \delta\rank(\ypart|\x) \cdot \relx(\ypart)  \right] > 0 \\
   \!\!& \Leftrightarrow & \!\!\!\Risk(\system_1) > \Risk(\system_2),
\end{eqnarray*}
where $\delta\rank(\ypart|\x)$ is short for $\rank(\ypart|\system_1(\x)) - \rank(\ypart|\system_2(\x))$ and we use the fact that $\epsilon_- \sum_{\ypart \in \ypres} \delta\rank(\ypart|\x) = 0$ in the step marked $*$.
This implies that our propensity-weighted ERM is a consistent approach for finding a ranking function with the best true $\Risk(\system)$,
\begin{eqnarray}
    \hat{\system} & = & \argmin_{\system \in \systems} \left\{\Risk(\system) \right\} \nonumber \\
                  & = & \argmin_{\system \in \systems} \left\{\E[\Riskhat_{IPS}(\system)] \right\},
\end{eqnarray}
even when the objective is corrupted by click noise as specified above.

\subsection{Propensity Estimation}
\label{sec:prop_est}
As the last step of defining the click propensity model, we need to address the question of how to estimate its parameters (i.e. the vector of examination probabilities $p_r$) for a particular search engine. The following shows that we can get estimates using data from a simple intervention similar to \cite{Wang/etal/16}, but without the strong negative impact of presenting uniformly random results to some users. This also relates to the Click@1 metric proposed by \cite{Chapelle/Zhang/09}.

First, note that it suffices to estimate the $p_r$ up to some positive multiplicative constant, since any such constant does not change how the IPS estimator \eqref{eq:riskips_posmodel} orders different systems. We therefore merely need to estimate how much $p_r$ changes relative to $p_k$ for some ``landmark'' rank $k$. This suggests the following experimental intervention for estimating $p_r$: before presenting the ranking to the user, swap the result at rank $k$ with the result at rank $r$. If we denote with $\ypart'$ the results originally in rank $k$, our click model before and after the intervention indicates that
\begin{eqnarray*}
  P(\click_i(\ypart')=1 | \mbox{no-swap}) &\!\!\!=\!\!\!& p_k \cdot P(\click_i(\ypart')=1|\exam_i(\ypart')=1) \\
  P(\click_i(\ypart')=1 | \mbox{swap-k-and-r}) &\!\!\!=\!\!\!& p_r \cdot P(\click_i(\ypart')=1|\exam_i(\ypart')=1)
\end{eqnarray*}
where
\begin{eqnarray*}
    && \!\!\!\!\!\!\!\!\!\!P(\click_i(\ypart')=1|\exam_i(\ypart')=1)  \\
    && =\!\!\!\sum_{v\in\{0,1\}} \!\!\!\! P(\click_i(\ypart')\!=\!1|\relx_i(\ypart')\!=\!v,\exam_i(\ypart')\!=\!1) \cdot P(\relx_i(\ypart')\!=\!v)
\end{eqnarray*}
is constant regardless of the intervention. This means that the clickthrough rates $P(\click_i(\ypart')=1 | \mbox{swap-k-and-r})$, which we can estimate from the intervention data, are proportional to the parameters $p_r$ for any $r$. By performing the swapping intervention between rank $k$ and all other ranks $r$, we can estimate all the $p_r$ parameters.

This swap-intervention experiment is of much lower impact than the uniform randomization proposed in \cite{Wang/etal/16} for a different propensity estimation problem, and careful consideration of which rank $k$ to choose can further reduce impact of the swap experiment. From a practical perspective, it may also be unnecessary to separately estimate $p_r$ for each rank. Instead, one may want to interpolate between estimates at well-chosen ranks and/or employ smoothing. Finally, note that the intervention only needs to be applied on a small subset of the data used for fitting the click propensity model, while the actual data used for training the ERM learning algorithm does not require any interventions.

\subsection{Alternative Feedback Propensity Models}

The click propensity model we define above is arguably one of the simplest models one can employ for propensity modeling in LTR, and there is broad scope for extensions.

First, one could extend the model by incorporating other biases, for example, trust bias \cite{Joachims/etal/07a} which affects perceived relevance of a result based on its position in the ranking. This can be captured by conditioning the click probabilities also on the position $\Prob(\click_i(\ypart')=1|\relx_i(\ypart'),\exam_i(\ypart')=1,\rank(\ypart|\ypres_i))$. We have already explored that the model can be extended to include trust bias, 
but it is omitted due to space constraints. Furthermore, it is possible to model saliency biases \cite{Yue2010a} by replacing the $p_r$ with a regression function.

Second, we conjecture that a wide range of other click models (e.g., cascade model \cite{craswell2008position} and others \cite{craswell2008position,Chapelle/Zhang/09,borisov2016neural,Chuklin/etal/15a}) can be adapted as propensity models. The main requirement is that we can compute marginal click probabilities for the clicked documents in hindsight, which is computationally feasible for many of the existing models.

Third, we may be able to define and train new types of click models. In particular, for our propensity ERM approach we only need the propensities $Q(\obs_{i}(\ypart)=1|\x_i,\ypres_i,\relx_i)$ for observed and relevant documents to evaluate the IPS estimator, but not for irrelevant documents. This can be substantially easier than a full generative model of how people reveal relevance judgments through implicit feedback. In particular, this model can condition on all the revealed relevances $\relx_i(\ypart_j)$ in hindsight, and it does not need to treat them as latent variables.

Finally, the ERM learning approach is not limited to binary click feedback, but applies to a large range of feedback settings. For example, the feedback may be explicit star ratings in a movie recommendation system, and the propensities may be the results of self-selection by the users as in \cite{Schnabel/etal/16b}. In such an explicit feedback setting, $\obs_i$ is fully known, which simplifies propensity estimation substantially.

\section{Propensity-weighted SVM-Rank} \label{sec:propensitysvm}

We now derive a concrete learning method that implements propensity-weighted LTR. It is based on SVM-Rank \cite{Joachims/02c,Joachims/06a}, but we conjecture that propensity-weighted versions of other 
LTR methods can be derived as well.

Consider a dataset of $n$ examples of the following form. For each query-result pair $(\x_j,\ypart_j)$ that is clicked, we compute the propensity $q_i=Q(\obs_{i}(\ypart)=1|\x_i,\ypres_i,\relx_i)$ of the click according to our click propensity model. We also record the candidate set $Y_j$ of all results for query $\x_j$. Typically, $Y_j$ contains a few hundred documents -- selected by a stage-one ranker \cite{Wang/etal/11a} -- that we aim to rerank. Note that each click generates a separate training example, even if multiple clicks occur for the same query.

Given this propensity-scored click data, we define Propensity SVM-Rank as a generalization of conventional SVM-Rank. Propensity SVM-Rank learns a linear scoring function $f(\x,\ypart)=\w \cdot \phi(\x,\ypart)$ that can be used for ranking results, where $\w$ is a weight vector and $\phi(\x,\ypart)$ is a feature vector that describes the match between query $\x$ and result $\ypart$.


Propensity SVM-Rank optimizes the following objective,
\begin{eqnarray*}
    \what & \!\!=\!\! & \argmin_{w,\xi} \: \frac{1}{2} \w \cdot \w + \frac{C}{n} \sum_{j=1}^n \frac{1}{q_j} \sum_{\ypart \in Y_j} \xi_{j\ypart} \\
     s.t. &   & \forall \ypart \in Y_1 \!\setminus\! \{\ypart_1\}: \w \cdot [\phi(\x_1,\ypart_1) - \phi(\x_1,\ypart)] \ge 1\!-\!\xi_{1\ypart} \\
          &   & \hspace{1cm} \vdots \\
          &   & \forall \ypart \in Y_n \!\setminus\! \{\ypart_n\}: \w \cdot [\phi(\x_n,\ypart_n) - \phi(\x_n,\ypart)] \ge 1\!-\!\xi_{n\ypart} \\
          &   & \forall j \forall \ypart: \xi_{j\ypart} \geq 0.
\end{eqnarray*}

$C$ is a regularization parameter that is typically selected via cross-validation. The training objective optimizes an upper bound on the regularized IPS estimated empirical risk of \eqref{eq:riskips_posmodel}, since each line of constraints corresponds to the rank of a relevant document (minus 1). In particular, for any feasible ($\w,\xi$)
\begin{align*}
    \rank(\ypart_i|\y) - 1 & =  \sum_{\ypart \not= \ypart_i} \mathbbm{1}_{\w \cdot [\phi(\x_i,\ypart) - \phi(\x_i,\ypart_i)] > 0}\\
    & \leq  \sum_{\ypart \not= \ypart_i} \max(1 - \w \cdot [\phi(\x_i,\ypart_i) - \phi(\x_i,\ypart)], 0) \\
    & \leq \sum_{\ypart \not= \ypart_i} \xi_{i\ypart}.
\end{align*}

\begin{sloppypar}
We can solve this type of Quadratic Program
efficiently via a one-slack formulation \cite{Joachims/06a}, and we are using SVM-Rank\footnote{https://www.joachims.org/svm\_light/svm\_rank.html} with appropriate modifications to include IPS weights $1/q_j$. The resulting code will be available online.
\end{sloppypar}

In the empirical evaluation, we compare against the naive application of SVM-Rank, which minimizes the rank of the clicked documents while ignoring presentation bias. In particular, Naive SVM-Rank sets all the $q_i$ uniformly to the same constant (e.g., $1$).

\section{Empirical Evaluation}

We take a two-pronged approach to evaluating our approach empirically. First, we use synthetically generated click data to explore the behavior of our methods over the whole spectrum of presentation bias severity, click noise, and propensity misspecification. Second, we explore the real-world applicability of our approach by evaluating on an operational search engine using real click-logs from live traffic. 

\subsection{Synthetic Data Experiments} \label{sec:synth_expt}

To be able to explore the full spectrum of biases and noise, we conducted experiments using click data derived from the Yahoo Learning to Rank Challenge corpus (set 1). This corpus contains a large number of manually judged queries, where we binarized relevance by assigning $\relx_i(\ypart)=1$ to all documents that got rated $3$ or $4$, and $\relx_i(\ypart)=0$ for ratings $0, 1, 2$. We adopt the train, validation, test splits in the corpus. This means that queries in the three sets are disjoint, and we never train on any data from queries in the test set. To have a gold standard for reporting test-set performance, we measure performance on the binarized full-information ratings using \eqref{eq:avgrankfullinfo}. 

To generate click data from this full-information dataset of ratings, we first trained a normal Ranking SVM using 1 percent of the full-information training data to get a ranking function $\system_0$. We employ $\system_0$ as the ``Production Ranker'', and it is used to ``present'' rankings $\ypres$ when generating the click data. We generate clicks using the rankings $\ypres$ and ground-truth binarized relevances from the Yahoo dataset according to the following process. Depending on whether we are generating a training or a validation sample of click data, we first randomly draw a query $\x$ from the respective full-information dataset. For this query we compute $\ypres=\system_0(\x)$ and generate clicks based on the model from Section~\ref{sec:feedbackmodels}. Whenever a click is generated, we record a training example with its associated propensity $Q(\obs(\ypart)=1|\x,\ypres,\relx)$. For the experiments, we model presentation bias via
\begin{eqnarray}
    Q(\obs(\ypart)=1|\x,\ypres,\relx) = p_{rank(\ypart|\ypres)} = \left(\frac{1}{rank(\ypart|\ypres)}\right)^\eta \label{eq:clickbias}.
\end{eqnarray}
The parameter $\eta$ lets us control the severity of the presentation bias. We also introduce noise into the clicks according to the model described in Section~\ref{sec:feedbackmodels}. When not mentioned otherwise, we use the parameters $\eta=1$, $\epsilon_-=0.1$, and $\epsilon_+=1$, which leads to click data where about $33\%$ of the clicks are noisy clicks on irrelevant results and where the result at rank $10$ has a $10\%$ probability of being examined. We also explore other bias profiles and noise levels in the following experiments.

In all experiments, we select any parameters (e.g., $C$) of the learning methods via cross-validation on a validation set. The validation set is generated using the same click model as the training set, but using the queries in the validation-set portion of the Yahoo dataset. For Propensity SVM-Rank, we always use the (unclipped) IPS estimator \eqref{eq:riskips_posmodel} to estimate validation set performance. Keeping with the proportions of the original Yahoo data, the validation set size is always about $15\%$ the size of the training set.

The primary baseline we compare against is a naive application of SVM-Rank that simply ignores the bias in the click data. We call this method {\em Naive SVM-Rank}. It is equivalent to a standard ranking SVM \cite{Joachims/02c}, but is most easily explained as equivalent to Propensity SVM-Rank with all $q_{j}$ set to $1$. Analogously, we use the corresponding naive version of \eqref{eq:riskips_posmodel} with propensities set to $1$ to estimate validation set performance for Naive SVM-Rank. 

\subsection{How does ranking performance scale with training set size?}

\begin{figure}
    \centering
    \includegraphics*[width=0.9\linewidth,trim={0.5cm 0cm 0.4cm 0.4cm},clip]{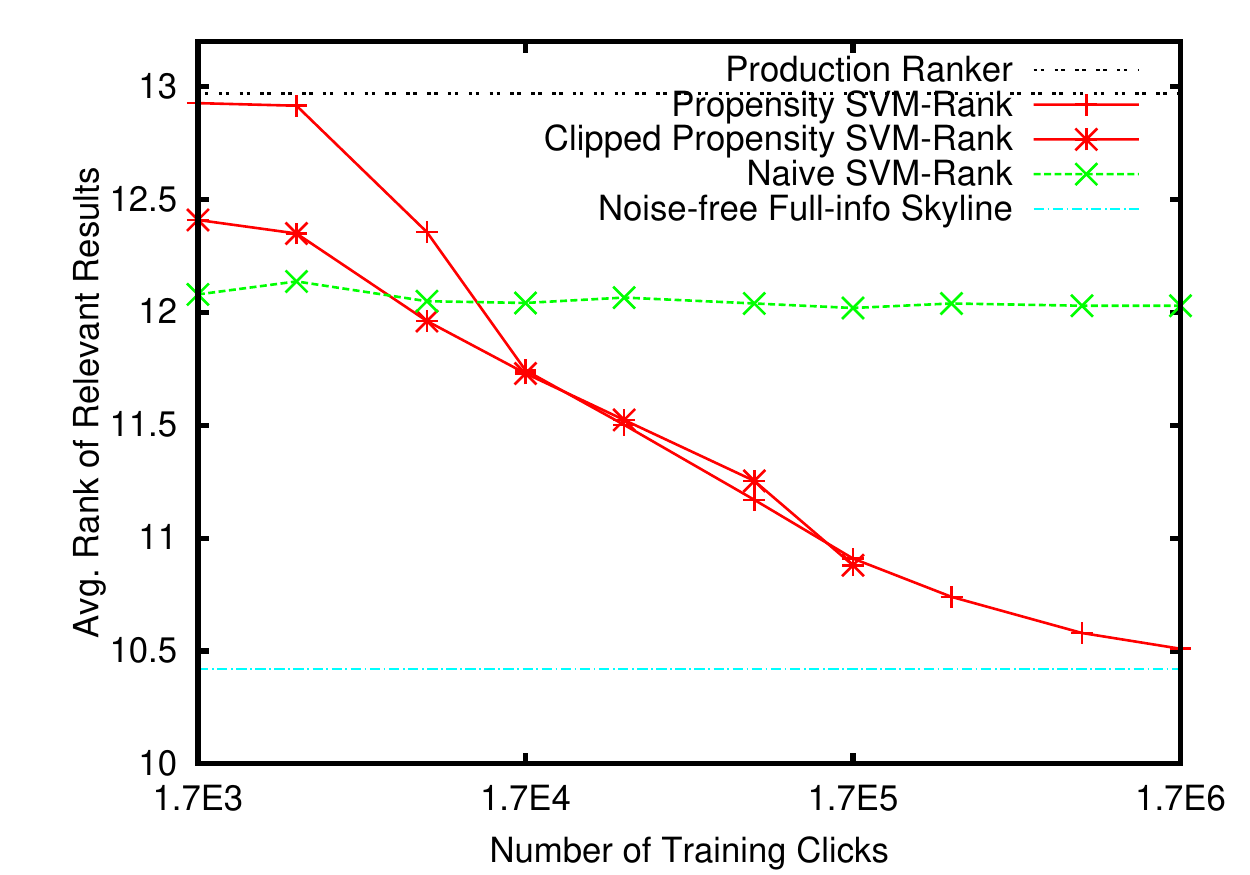}
    \vspace*{-0.3cm}
    \caption{Test set performance in terms of \eqref{eq:avgrankfullinfo} for Propensity SVM-Rank with and without clipping compared to SVM-Rank naively ignoring the bias in clicks ($\eta=1$, $\epsilon_-=0.1$). The skyline is a Ranking SVM trained on all data without noise in the full-information setting, and the baseline is the production ranker $\system_0$.}
    \label{fig:exp_n}
\end{figure}

We first explore how the test-set ranking performance changes as the learning algorithm is given more and more click data. The resulting learning curves are given in Figure~\ref{fig:exp_n}, and the performance of $\system_0$ is given as a baseline. The click data has presentation bias according to \eqref{eq:avgrankfullinfo} with $\eta=1$ and noise $\epsilon_-=0.1$. For small datasets, results are averaged over 5 draws of the click data.

\begin{sloppypar}
With increasing amounts of click data, Propensity SVM-Rank approaches the skyline performance of the full-information SVM-Rank trained on the complete training set of manual ratings without noise. This is in stark contrast to Naive SVM-Rank which fails to account for the bias in the data and does not reach this level of performance. Furthermore, Naive SVM-Rank cannot make effective use of additional data and its learning curve is essentially flat. This is consistent with the theoretical insight that estimation error in Naive SVM-Rank's empirical risk $\Riskhat(\system)$ is dominated by asymptotic bias due to biased clicks, which does not decrease with more data and leads to suboptimal learning. The unbiased risk estimate $\Riskhat_{IPS}(\system)$ of Propensity SVM-Rank, however, has estimation error only due to finite sample variance, which is decreased by more data and leads to consistent learning.
\end{sloppypar}

While unbiasedness is an important property when click data is plenty, the increased variance of $\Riskhat_{IPS}(\system)$ can be a drawback for small datasets. This can be seen in Figure~\ref{fig:exp_n}, where Naive SVM-Rank outperforms Propensity SVM-Rank for small datasets. This can be remedied using techniques like ``propensity clipping'' \cite{Strehl2010}, where small propensities are clipped to some threshold value $\tau$ to trade bias for variance.
\begin{eqnarray*}
    \Riskhat_{CIPS}(\system) & \!\!\!=\!\!\! & \frac{1}{\n}\sum_{\x_i} \sum_{\ypart \in \system(\x_i)} \!\!\!\frac{\rank(\ypart|\system(\x_i)) \cdot \relx_i(\ypart)}{\max\{ \tau, Q(\obs_{i}(\ypart)\!=\!1|\x_i,\ypres_i, \relx_i)\}}. 
\end{eqnarray*}
Figure~\ref{fig:exp_n} shows the learning curve of Propensity SVM-Rank with clipping, cross-validating both the clipping threshold $\tau$ and $C$. Clipping indeed improves performance for small datasets. While $\tau=1$ is equivalent to Naive SVM-Rank, the validation set is too small (and hence, the finite sample error of the validation performance estimate too high) to reliably select this model in every run. In practice, however, we expect click data to be plentiful such that lack of training data is unlikely to be a persistent issue.

\subsection{How much presentation bias can be tolerated?}

\begin{figure}
    \centering
    \includegraphics*[width=0.9\linewidth,trim={0.5cm 0cm 0.4cm 0.4cm},clip]{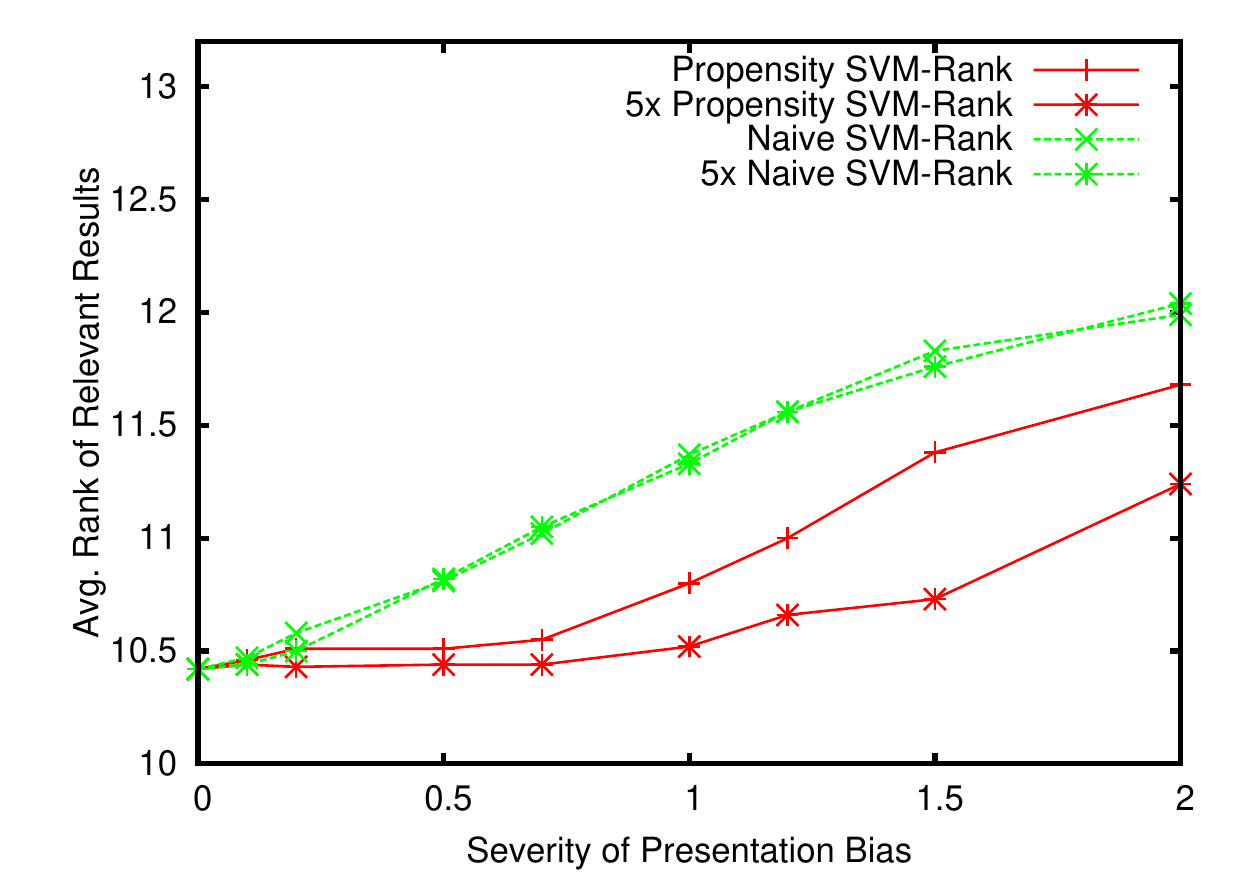}
    \vspace*{-0.3cm}
    \caption{Test set performance for Propensity SVM-Rank and Naive SVM-Rank as presentation bias becomes more severe in terms of $\eta$ ($n=45K$ and $n=225K$, $\epsilon_-=0$).}
    \label{fig:exp_prop}
\end{figure}

We now vary the severity of the presentation bias via $\eta$ to understand its impact on Propensity SVM-Rank. Figure~\ref{fig:exp_prop} shows that inverse propensity weighting is beneficial whenever substantial bias exists. Furthermore, increasing the amount of training data by a factor of $5$ leads to further improvement for the Propensity SVM-Rank, while the added training data has no effect on Naive SVM-Rank. This is consistent with our arguments from Section~\ref{sec:partial_ltr} -- more training data does not help when bias dominates estimation error, but it can reduce estimation error from variance in the unbiased risk estimate of Propensity SVM-Rank.

\subsection{How robust are the methods to click noise?}

\begin{figure}
    \centering
    \includegraphics*[width=0.9\linewidth,trim={0.5cm 0cm 0.4cm 0.4cm},clip]{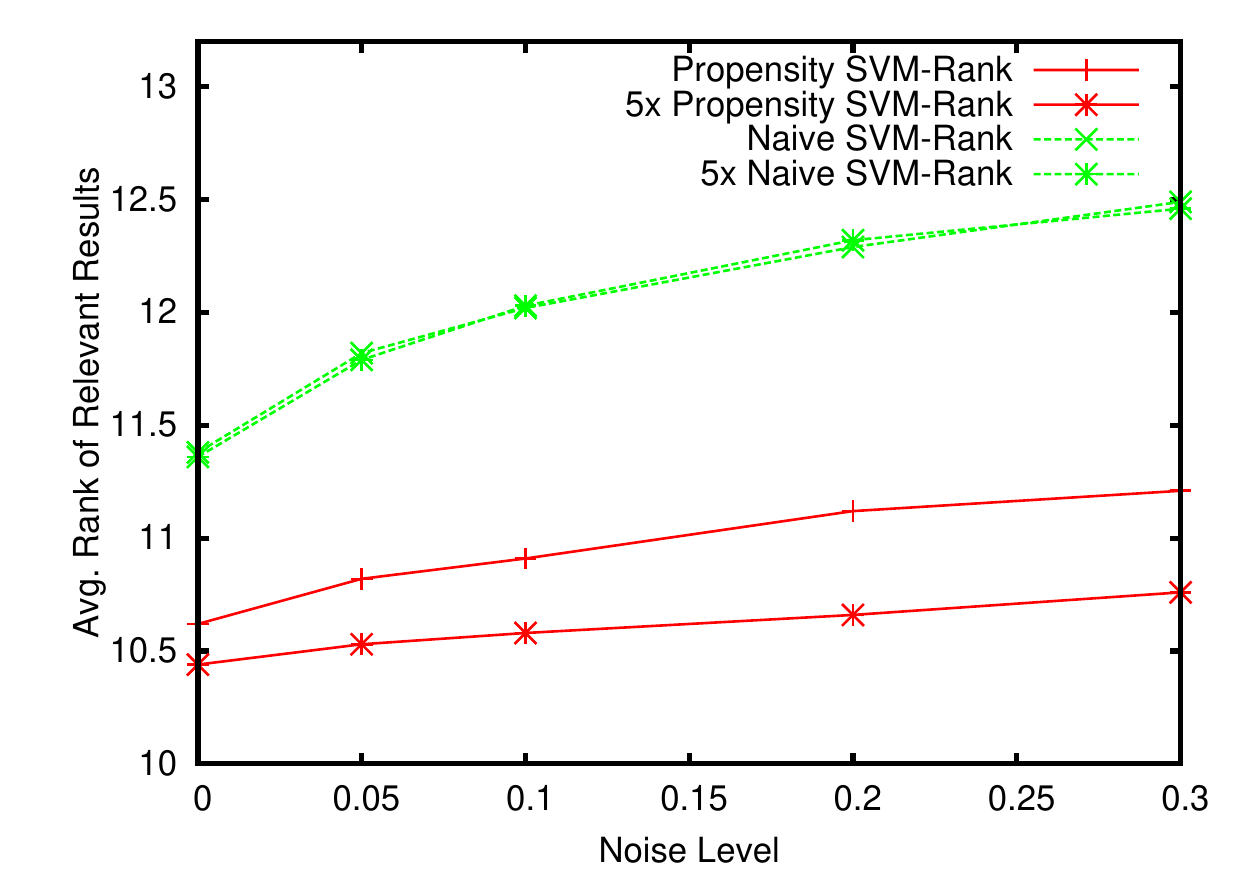}
    \vspace*{-0.3cm}
    \caption{Test set performance for Propensity SVM-Rank and Naive SVM-Rank as the noise level increases in terms of $\epsilon_-$ ($n=170K$ and $n=850K$, $\eta=1$).}
    \label{fig:exp_noise}
\end{figure}

Figure~\ref{fig:exp_noise} shows that Propensity SVM-Rank also enjoys a substantial advantage when it comes to noise. When increasing the noise level in terms of $\epsilon_-$ from $0$ up to $0.3$ (resulting in click data where $59.8\%$ of all clicks are on irrelevant documents), Propensity SVM-Rank increasingly outperforms Naive SVM-Rank. And, again, the unbiasedness of the empirical risk estimate allows Propensity SVM-Rank to benefit from more data. 


\subsection{How robust is Propensity SVM-Rank to misspecified propensities?}

\begin{figure}
    \centering
    \includegraphics*[width=0.9\linewidth,trim={0.5cm 0cm 0.4cm 0.4cm},clip]{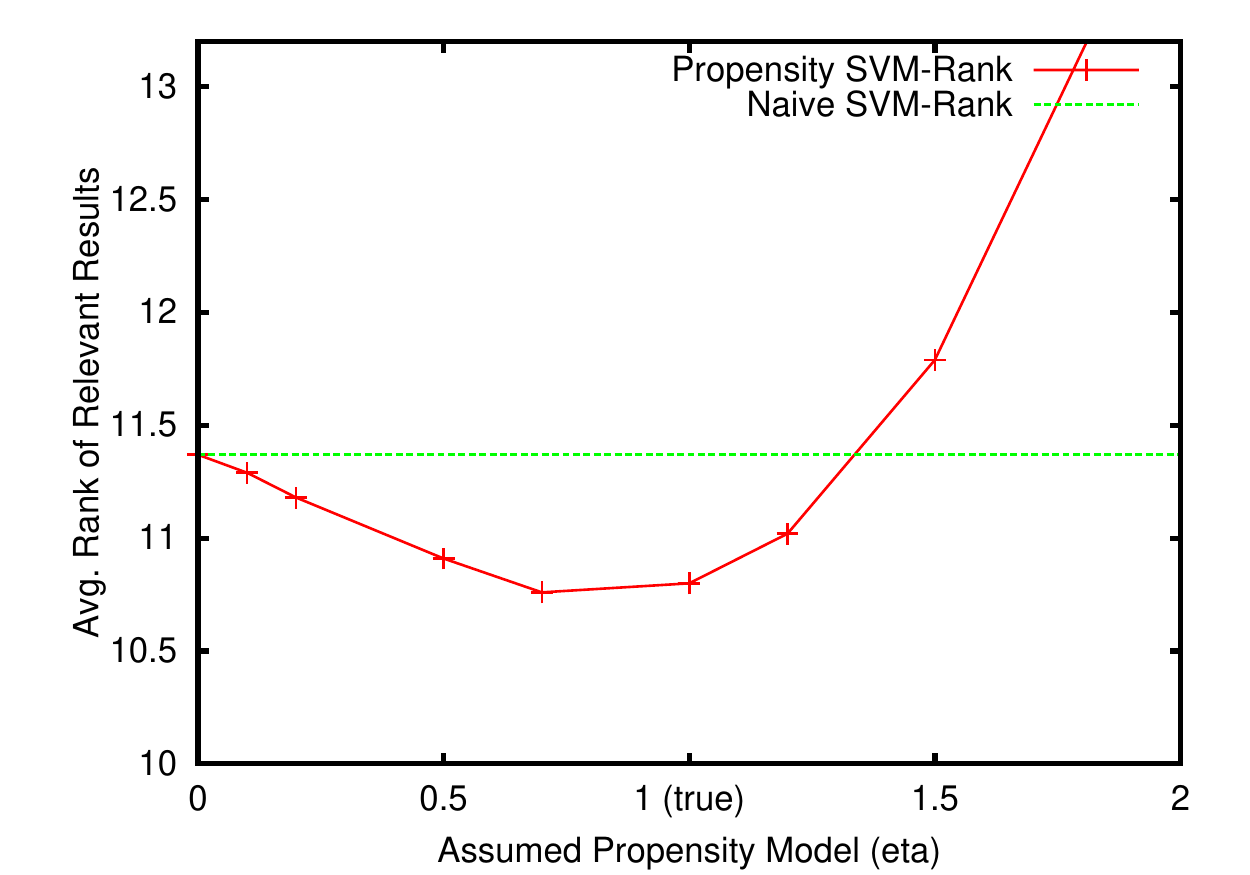}
    \vspace*{-0.3cm}
    \caption{Test set performance for Propensity SVM-Rank and Naive SVM-Rank as propensities are misspecified (true $\eta=1$, $\n=170K$, $\epsilon_-=0.1$).}
    \label{fig:exp_mismatch}
\end{figure}

So far all experiments have assumed that Propensity SVM-Rank has access to accurate propensities. In practice, however, propensities need to be estimated and are subject to model assumptions. We now evaluate how robust Propensity SVM-Rank is to misspecified propensities. Figure~\ref{fig:exp_mismatch} shows the performance of Propensity SVM-Rank when the training data is generated with $\eta=1$, but the propensities used by Propensity SVM-Rank are misspecified using the $\eta$ given in the x-axis of the plot. The plot shows that even misspecified propensities can give substantial improvement over naively ignoring the bias, as long as the misspecification is ``conservative'' -- i.e., overestimating small propensities is tolerable (which happens when $\eta<1$), but underestimating small propensities can be harmful (which happens when $\eta>1$). This is consistent with theory, and clipping is one particular way of overestimating small propensities that can even improve performance. Overall, we conclude that even a mediocre propensity model can improve over the naive approach -- after all, the naive approach can be thought of as a particularly poor propensity model that implicitly assumes no presentation bias and uniform propensities.

\subsection{Real-World Experiment}

We now examine the performance of Propensity SVM-rank when trained on real-world click logs and deployed in a live search engine for scientific articles [anonymized for submission].
The search engine uses a linear scoring function as outlined in Section~\ref{sec:propensitysvm}. Query-document features $\phi(\x,\ypart)$ are represented by a $1000-$dimensional vector, and the production ranker 
used for collecting training clicks employs a hand-crafted weight vector $w$ (denoted Prod).
Observed clicks on rankings served by this ranker over a period of $21$ days provide implicit feedback data for LTR as outlined in Section~\ref{sec:propensitysvm}.

To estimate the propensity model, 
we consider the simple position-based model of Section~\ref{sec:position_feedbackmodel} and
we collect new click data via randomized interventions for $7$ days as outlined in Section~\ref{sec:prop_est} with landmark rank $k=1$. 
Before presenting the ranking, we take the top-ranked document and swap it with the document at a uniformly at random chosen rank $j \in \{ 1,\dots  21\}$.  
The ratio of observed click-through rates (CTR) on the formerly top-ranked document now at position $j$ vs. its CTR at position $1$ gives a noisy estimate of $p_j/p_1$ in the position-based click model. We additionally smooth these estimates by interpolating with the overall observed CTR at position $j$ (normalized so that $CTR@1 = 1$). This yields $p_r$ that approximately decay with rank $r$ with the smallest $p_r \simeq 0.12$. For $r>21$, we impute $p_r=p_{21}$.

We partition the click-logs into a train-validation split: the first $16$ days are the train set and provide $5437$ click-events for SVM-rank, while the remaining $5$ days are the validation set with $1755$ click events. 
The hyper-parameter $C$ is picked via cross validation. Analogous to Section~\ref{sec:synth_expt}, we use the IPS estimator for Propensity SVM-Rank, and naive estimator with $Q(\obs(\ypart)=1|\x,\ypres,\relx)=1$ for Naive SVM-Rank. With the best hyper-parameter settings, we re-train on all $21$ days worth of data to derive the final weight vectors for either method.

\begin{table}[t]
\vspace{-0.15in}
\caption{Per-query balanced interleaving results for detecting relative performance between the hand-crafted production ranker used for click data collection (Prod), Naive SVM-Rank and Propensity SVM-Rank. }
\label{tab:arxiv_expt}
\vspace{-0.08in}
\begin{center}
\begin{tabular}{|l|c|c|c|}
\hline
           & \multicolumn{3}{c|}{Propensity SVM-Rank}  \\
Interleaving Experiment & $\:\:$ wins $\:\:$ & $\:\:$ loses $\:\:$ & $\:\:$ ties $\:\:$\\
\hline
against Prod & 87 & 48 & 83 \\
against Naive SVM-Rank & 95 & 60 & 102 \\
\hline
\end{tabular}
\end{center}
\vspace{-0.2in}
\end{table}

We fielded these learnt weight vectors in two online interleaving experiments \cite{Chapelle/etal/12a}, the first comparing Propensity SVM-Rank against Prod 
and the second comparing Propensity SVM-Rank against Naive SVM-Rank.
The results are summarized in Table~\ref{tab:arxiv_expt}. We find that Propensity SVM-Rank significantly outperforms the hand-crafted production ranker that was used to collect the click data for training (two-tailed binomial sign test $p=0.001$ with relative risk $0.71$ compared to null hypothesis). Furthermore, Propensity SVM-Rank similarly outperforms Naive SVM-Rank, demonstrating that even a simple propensity model provides benefits on real-world data (two-tailed binomial sign test $p=0.006$ with relative risk $0.77$ compared to null hypothesis). 
Note that Propensity SVM-Rank not only significantly, but also substantially outperforms both other rankers in terms of effect size -- and the synthetic data experiments suggest that additional training data will further increase its  advantage.

\section{Conclusions}

This paper introduced a principled approach for learning-to-rank under biased feedback data. Drawing on counterfactual modeling techniques from causal inference, we present a theoretically sound Empirical Risk Minimization framework for LTR.  We instantiate this framework with a Propensity-Weighted Ranking SVM, and provide extensive empirical evidence that the resulting learning method is robust to selection biases, noise, and model misspecification. Furthermore, our real-world experiments on a live search engine show that the approach leads to substantial retrieval improvements,  without any heuristic or manual interventions in the learning process. 

\section{Future Research}

 Beyond the specific learning methods and propensity models we propose, this paper may have even bigger impact for its theoretical contribution of developing the general counterfactual model for LTR, thus articulating the key components necessary for LTR under biased feedback. First, the insight that propensity estimates are crucial for ERM learning opens a wide area of research on designing better propensity models. Second, the theory demonstrates that LTR methods should optimize propensity-weighted ERM objectives, raising the question of which other learning methods beyond the Ranking SVM can be adapted to the Propensity ERM approach. Third, we conjecture that a Propensity ERM approach can be developed also for pointwise LTR methods using techniques from \cite{Schnabel/etal/16c}, and possibly even for listwise LTR.

\begin{sloppypar}
Beyond learning from implicit feedback, propensity-weighted ERM techniques may prove useful even for optimizing offline IR metrics on manually annotated test collections. First, they can eliminate pooling bias, since the use of sampling during judgment elicitation puts us in a controlled setting where propensities are known (and can be optimized \cite{Schnabel/etal/16c}) by design. Second, propensities estimated via click models can enable click-based IR metrics like click-DCG to better correlate with test set DCG.
\end{sloppypar}

\iftoggle{ARXIV}{
\pagebreak[4]
\section{Acknowledgments}
This work was supported in part through NSF Awards IIS-1247637, IIS-1513692, IIS-1615706, and a gift from Bloomberg. We thank Maarten de Rijke, Alexey Borisov, Artem Grotov, and Yuning Mao for valuable feedback and discussions.
}{
\pagebreak[4]
}
\bibliographystyle{abbrv}
\bibliography{references,joachims}

\begin{thebibliography}{10}

\bibitem{borisov2016neural}
A.~Borisov, I.~Markov, M.~de~Rijke, and P.~Serdyukov.
\newblock A neural click model for web search.
\newblock In {\em Proceedings of the 25th International Conference on World
  Wide Web}, pages 531--541, 2016.

\bibitem{Chapelle/etal/12a}
O.~Chapelle, T.~Joachims, F.~Radlinski, and Y.~Yue.
\newblock Large-scale validation and analysis of interleaved search evaluation.
\newblock {\em ACM Transactions on Information Systems (TOIS)},
  30(1):6:1--6:41, 2012.

\bibitem{Chapelle/Zhang/09}
O.~Chapelle and Y.~Zhang.
\newblock A dynamic bayesian network click model for web search ranking.
\newblock In {\em International Conference on World Wide Web (WWW)}, pages
  1--10. ACM, 2009.

\bibitem{Chuklin/etal/15a}
A.~Chuklin, I.~Markov, and M.~de~Rijke.
\newblock {\em Click Models for Web Search}.
\newblock Synthesis Lectures on Information Concepts, Retrieval, and Services.
  Morgan \& Claypool Publishers, 2015.

\bibitem{craswell2008position}
N.~Craswell, O.~Zoeter, M.~Taylor, and B.~Ramsey.
\newblock An experimental comparison of click position-bias models.
\newblock In {\em International Conference on Web Search and Data Mining
  (WSDM)}, pages 87--94. ACM, 2008.

\bibitem{hofmann2013reusing}
K.~Hofmann, A.~Schuth, S.~Whiteson, and M.~de~Rijke.
\newblock Reusing historical interaction data for faster online learning to
  rank for ir.
\newblock In {\em International Conference on Web Search and Data Mining
  (WSDM)}, pages 183--192, 2013.

\bibitem{Horvitz1952}
D.~G. Horvitz and D.~J. Thompson.
\newblock A generalization of sampling without replacement from a finite
  universe.
\newblock {\em Journal of the American Statistical Association},
  47(260):663--685, 1952.

\bibitem{Imbens/Rubin/15}
G.~Imbens and D.~Rubin.
\newblock {\em Causal Inference for Statistics, Social, and Biomedical
  Sciences}.
\newblock Cambridge University Press, 2015.

\bibitem{Joachims/02c}
T.~Joachims.
\newblock Optimizing search engines using clickthrough data.
\newblock In {\em ACM SIGKDD Conference on Knowledge Discovery and Data Mining
  (KDD)}, pages 133--142, 2002.

\bibitem{Joachims/06a}
T.~Joachims.
\newblock Training linear {SVMs} in linear time.
\newblock In {\em ACM SIGKDD International Conference On Knowledge Discovery
  and Data Mining (KDD)}, pages 217--226, 2006.

\bibitem{Joachims/etal/07a}
T.~Joachims, L.~Granka, B.~Pan, H.~Hembrooke, F.~Radlinski, and G.~Gay.
\newblock Evaluating the accuracy of implicit feedback from clicks and query
  reformulations in web search.
\newblock {\em ACM Transactions on Information Systems (TOIS)}, 25(2), April
  2007.

\bibitem{li2011unbiased}
L.~Li, W.~Chu, J.~Langford, and X.~Wang.
\newblock Unbiased offline evaluation of contextual-bandit-based news article
  recommendation algorithms.
\newblock In {\em International Conference on Web Search and Data Mining
  (WSDM)}, pages 297--306, 2011.

\bibitem{rubin2002mnar}
R.~J.~A. Little and D.~B. Rubin.
\newblock {\em Statistical Analysis with Missing Data}.
\newblock John Wiley, 2002.

\bibitem{Liu2009}
T.-Y. Liu.
\newblock Learning to rank for information retrieval.
\newblock {\em Foundations and Trends in Information Retrieval}, 3(3):225--331,
  Mar. 2009.

\bibitem{Raman/Joachims/13a}
K.~Raman and T.~Joachims.
\newblock Learning socially optimal information systems from egoistic users.
\newblock In {\em European Conference on Machine Learning (ECML)}, pages
  128--144, 2013.

\bibitem{Raman/etal/13a}
K.~Raman, T.~Joachims, P.~Shivaswamy, and T.~Schnabel.
\newblock Stable coactive learning via perturbation.
\newblock In {\em International Conference on Machine Learning (ICML)}, pages
  837--845, 2013.

\bibitem{Richardson2007}
M.~Richardson, E.~Dominowska, and R.~Ragno.
\newblock Predicting clicks: Estimating the click-through rate for new ads.
\newblock In {\em International Conference on World Wide Web (WWW)}, pages
  521--530. ACM, 2007.

\bibitem{Rosenbaum1983}
P.~R. Rosenbaum and D.~B. Rubin.
\newblock The central role of the propensity score in observational studies for
  causal effects.
\newblock {\em Biometrika}, 70(1):41--55, 1983.

\bibitem{Schnabel/etal/16c}
T.~Schnabel, A.~Swaminathan, P.~Frazier, and T.~Joachims.
\newblock Unbiased comparative evaluation of ranking functions.
\newblock In {\em ACM International Conference on the Theory of Information
  Retrieval (ICTIR)}, 2016.

\bibitem{Schnabel/etal/16b}
T.~Schnabel, A.~Swaminathan, A.~Singh, N.~Chandak, and T.~Joachims.
\newblock Recommendations as treatments: Debiasing learning and evaluation.
\newblock In {\em International Conference on Machine Learning (ICML)}, 2016.

\bibitem{schuth2016multileave}
A.~Schuth, H.~Oosterhuis, S.~Whiteson, and M.~de~Rijke.
\newblock Multileave gradient descent for fast online learning to rank.
\newblock In {\em International Conference on Web Search and Data Mining
  (WSDM)}, pages 457--466, 2016.

\bibitem{SparckJones1975}
K.~Sparck-Jones and C.~J.~V. Rijsbergen.
\newblock Report on the need for and provision of an “ideal” information
  retrieval test collection.
\newblock Technical report, University of Cambridge, 1975.

\bibitem{Strehl2010}
A.~L. Strehl, J.~Langford, L.~Li, and S.~Kakade.
\newblock Learning from logged implicit exploration data.
\newblock In {\em Proceedings of the 24th Annual Conference on Neural
  Information Processing Systems}, pages 2217--2225, 2010.

\bibitem{Swaminathan/Joachims/15c}
A.~Swaminathan and T.~Joachims.
\newblock Batch learning from logged bandit feedback through counterfactual
  risk minimization.
\newblock {\em Journal of Machine Learning Research (JMLR)}, 16:1731--1755, Sep
  2015.
\newblock Special Issue in Memory of Alexey Chervonenkis.

\bibitem{Vapnik1998}
V.~Vapnik.
\newblock {\em Statistical Learning Theory}.
\newblock Wiley, Chichester, GB, 1998.

\bibitem{Wang/etal/11a}
L.~Wang, J.~J. Lin, and D.~Metzler.
\newblock A cascade ranking model for efficient ranked retrieval.
\newblock In {\em {ACM} Conference on Research and Development in Information
  Retrieval (SIGIR)}, pages 105--114. ACM, 2011.

\bibitem{Wang/etal/16}
X.~Wang, M.~Bendersky, D.~Metzler, and M.~Najork.
\newblock Learning to rank with selection bias in personal search.
\newblock In {\em ACM Conference on Research and Development in Information
  Retrieval (SIGIR)}. ACM, 2016.

\bibitem{Wang2016}
Y.~Wang, D.~Yin, L.~Jie, P.~Wang, M.~Yamada, Y.~Chang, and Q.~Mei.
\newblock Beyond ranking: Optimizing whole-page presentation.
\newblock In {\em Proceedings of the Ninth ACM International Conference on Web
  Search and Data Mining}, WSDM '16, pages 103--112, 2016.

\bibitem{Yue/Joachims/09a}
Y.~Yue and T.~Joachims.
\newblock Interactively optimizing information retrieval systems as a dueling
  bandits problem.
\newblock In {\em International Conference on Machine Learning (ICML)}, pages
  151--159, 2009.

\bibitem{Yue2010a}
Y.~Yue, R.~Patel, and H.~Roehrig.
\newblock Beyond position bias: examining result attractiveness as a source of
  presentation bias in clickthrough data.
\newblock In {\em International Conference on World Wide Web (WWW)}, pages
  1011--1018. ACM, 2010.

\end{thebibliography}

\end{document}